# Soft and Hard Interactions in proton + proton collisions at √s=200GeV

**By Shengli Huang (USTC) for STAR Collaboration**

Hadron interactions are often classified into “hard” and “soft” [1, 2] with jet as one of the characteristics of “hard” interactions. Previous results from CDF at collision energies 630~1800GeV observed that in the soft events the $<P_T>$ depends on multiplicity while not collision energies [3]. In this paper, we further check this independence in soft events at RHIC energy (200GeV) and study the baryon and strangeness production difference in the “soft” and “hard” interactions. These studies will help us understand the hadron interactions in p + p and more complicated Au + Au collisions.

There are 13 million non-singly diffractive proton + proton collisions collected by the STAR detector at RHIC during run 2. After offline analysis selection, 2.4 million events left for the inclusive charged hadron analysis and 3.6M for the identified particles analysis. We adapt the jet algorithm from CDF to separate “soft” and “hard” events [3]. The jet cluster is defined as at least two charged tracks in a cone with $R=\sqrt{\Delta\eta^2+\Delta\phi^2}=0.7$ in the full TPC η cover range ±1.4. The leading track is required $P_T>1.0$GeV/c, and the other track $P_T>0.2$ GeV/c. We define soft event as one that doesn’t contain such a jet cluster and the hard event as one containing at least one jet cluster. For the charged hadron analysis, the soft and hard events are about 1.8M and 0.6M. For the identified particles analysis, the soft and hard events are about 2.7M and 0.9M.

In STAR, the efficiency and acceptance of TPC are obtained by embedding the Monte Carlo (MC) tracks into real data events. We also embed MC events into events which have empty triggers (abort-gap events) to study the background and fake of vertex reconstruction (fake vertex). To identified particle analysis, we use the Z= log (*dE/dx* / <*dE/dx*>) to extract the raw yield of $\pi^-$, $K^-$, $\bar{P}$ by four Gaussian functions fitting. The background of $\pi^-$ spectra from weak decay are about 12%, the $K^-$ and $\bar{P}$ background is negligible in the identified $P_T$ range. The point-to-point systematic uncertainties are less than 4% for $\pi^-$, $K^-$ and $\bar{P}$. For $K^-$ the point-point systematic errors are less than 12% in the $P_T$ bins significant overlap in *dE/dx* with electron. A correlated systematic uncertainty of 5% is estimated for all spectra.

Here we report the $<P_T>$ dependence on multiplicity at collision energy of 200GeV in three events classes and compare them with the results from CDF. For the pseudo-rapidity acceptance difference between two detectors, we used HIJING [7] to extend the acceptance of STAR from |η|<0.5 to |η|<1.0. The $<P_T>$ is extracted from the power law function fit to the invariant yield $P_T$ distribution spectra and integrating from 0.4GeV/c to ∝. As depicted in Fig1,

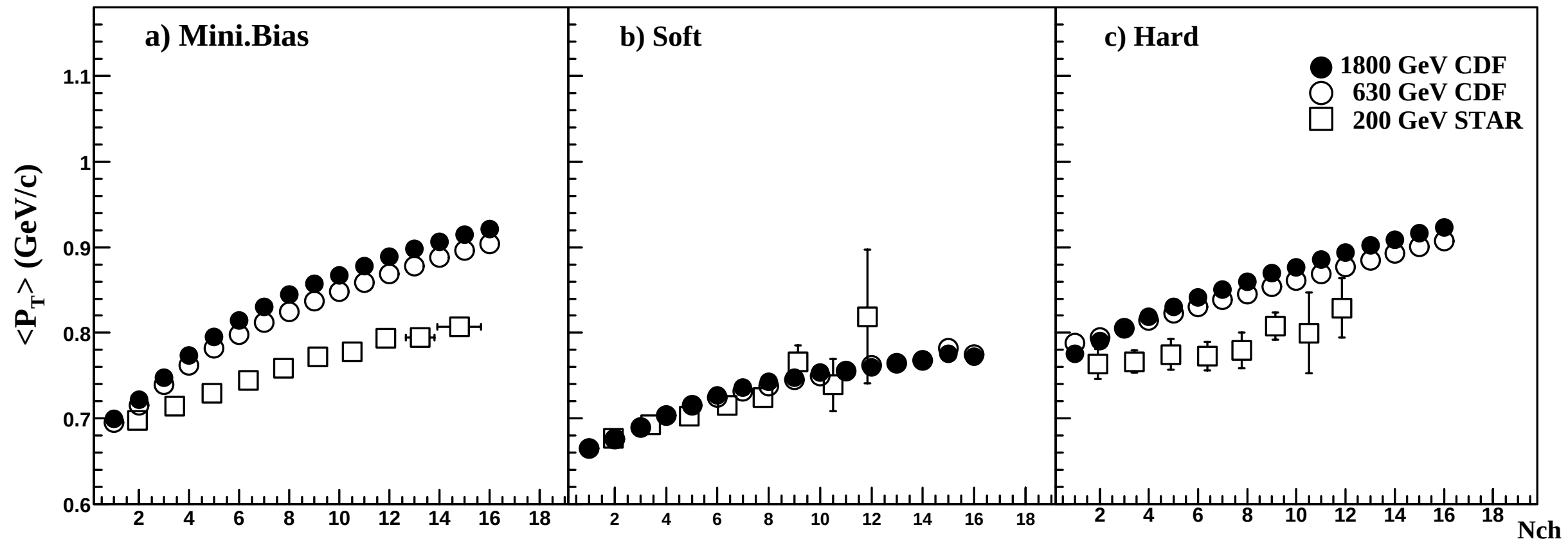

Fig1. The mean transverse momentum $<P_T>$ dependence on the multiplicity Nch in minimum bias, soft and hard events. The errors bars are statistical errors only. The STAR acceptance is extended from |η|<0.5 to |η|<1.0 by HIJING to compare the results with CDF.

the $<P_T>$ dependence on multiplicity in soft events are independent of collision energy from RHIC to Tevetron, but the minimum bias and hard events shows stronger dependence. These results confirm that the dynamical mechanism of inelastic multi-particle production in soft interactions from RHIC to Tevetron energy, is invariant with center of mass energy, and the properties of the final state are determined only by the number of (charged) particles. These can't be explained by current theoretical or phenomenological models [3]. In the soft event, the $<P_T>$ increases even in the small multiplicity, which indicates the mini-jet is not the only role of $<P_T>$ increasing with multiplicity [3].

The identified particle spectra from jet fragmentation have been studied in electron + positron collisions [8]. The study of identified particle spectra in different event classes in a simple proton + proton collision system can help us further understand the particle production in hard (semi-hard) and soft processes, and the experimental results from a more complex Au + Au collisions [9, 10, 11].

Due to the limited statistics, we divide the multiplicity into four bins as $N_{ch}$=1~3, 4~5, 6~7 and 9~13. After the corrections of background, efficiency, and fake vertex, the multiplicity densities are $dN_{ch}/d\eta=2.47\pm0.17$, $6.12\pm0.21$, $8.81 \pm 0.20$ and $12.41\pm0.20$, respectively. In order to characterize the shape of the spectra quantitatively, we fit the spectra with different functions. The $\pi^-$ spectra are fitted by three functions: (a) Bose-Einstein distribution in $m_T$ [$\propto 1/(\exp(m_T/T)-1)$], (b) exponential distribution in $m_T$ [$\propto \exp(-(m_T-m_0)/T)$] and exponential distribution in $P_T$ [$\propto \exp(-P_T)$]. The $K^-$ spectra are fitted by three functions as: (a) Bose-Einstein distribution in $m_T$, (b) exponential distribution in $m_T$ and (c) Maxwell-Boltzmann distribution in $m_T$ [$\propto m_T \exp(-(m_T-m_0)/T)$]. The $\bar{P}$ spectra are fitted by three functions as: (a) Gaussian distribution in $P_T$ [$\propto \exp(-P_T^2/2\sigma^2)$], (b) exponential distribution in $m_T$ and (c) Maxwell-Boltzmann distribution in $m_T$ [$\propto m_T \exp(-(m_T-m_0)/T)$]. These functions fit spectra equally well, with similar $\chi^2$ per degree of freedom of about 1. We can obtain the $<P_T>$ and total yield from these fit functions. The uncertainties on these extrapolations are estimated by comparing the fit results of three fit functions.

The multiplicity dependence of the extracted $<P_T>$ with $|y|<0.2$ is shown in Fig2. The systematic errors of $\pi^-$ are about 3% in minimum bias, soft events and hard events. The systematic errors for $K^-$, $\bar{P}$ are about 3%~7% in minimum bias events, 4% ~7% in soft events and 5%~9% in hard events from low multiplicity to high. The $<P_T>$ is almost flat as a function of multiplicity for $\pi^{--}$ in all three event classes. The $<P_T>$ increases with multiplicity to $K^-$, $\bar{P}$ even in soft events. This further indicates the mini-jet is not the only source responsible for the $<P_T>$ increase with multiplicity.

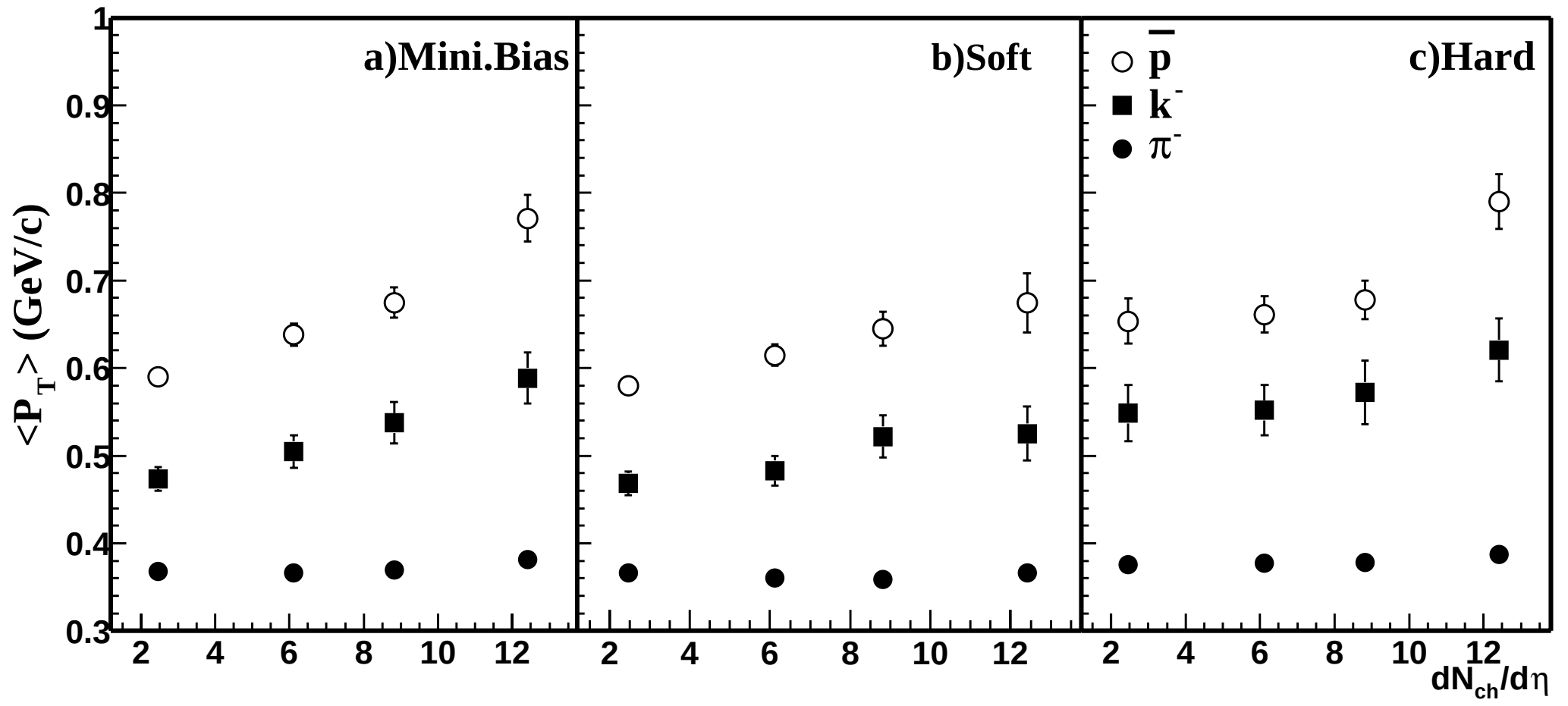

Fig2. The multiplicity dependence of the extracted <PT> with |y|<0.2.The error bars are statistic errors only. The systematic errors of π- are about 3% in minimum bias, soft events and hard events. The systematic errors for K-, $\bar{P}$ are about 3%~7% in minimum bias events, 4% in soft events and 5%~9% in hard events from low multiplicity to high.

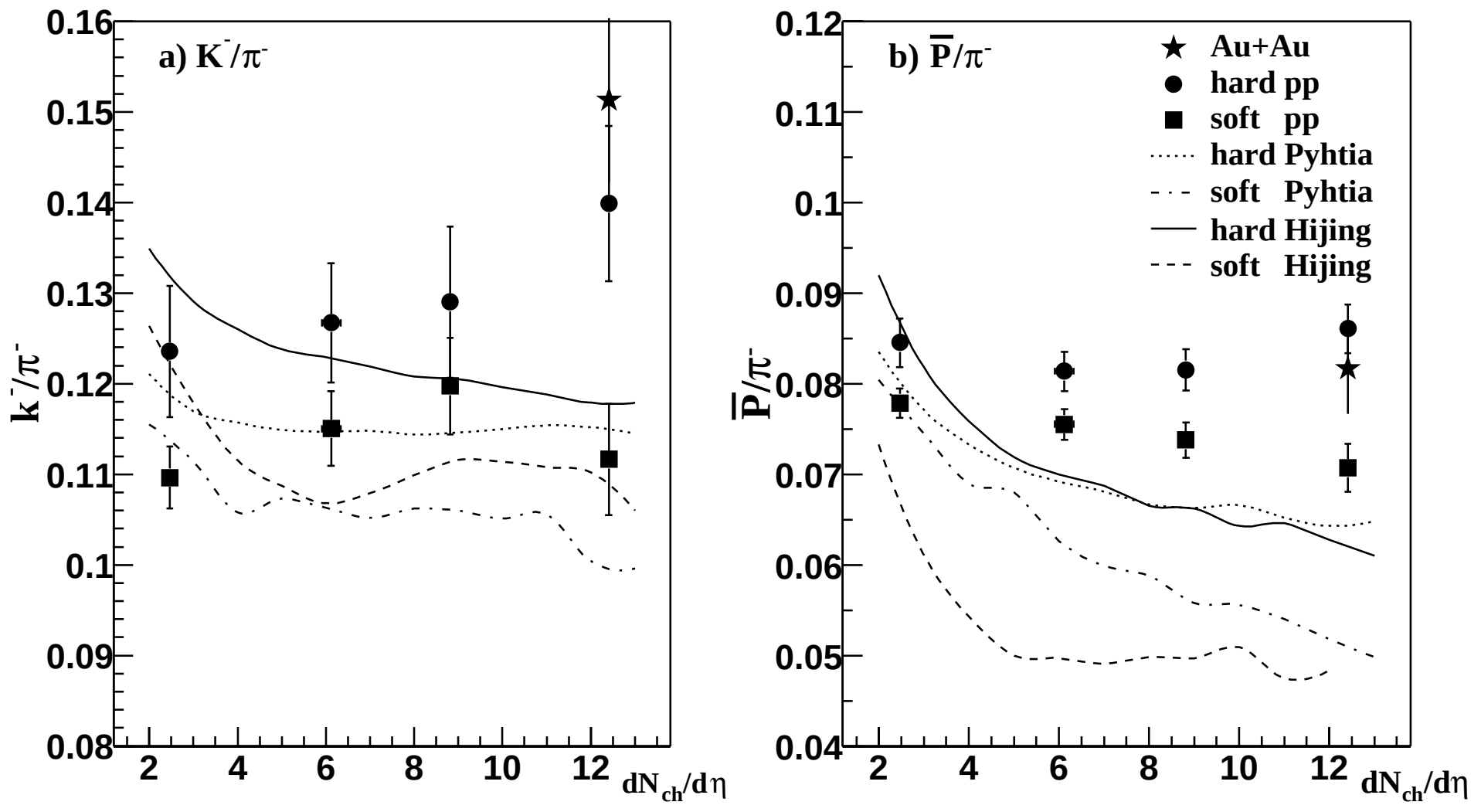

Fig3. The $K^-/\pi^-$, $\bar{P}$ $/\pi^-$ yield ratios in mid-rapidity (|y|<0.2).The yields are gotten by $m_T$ exponential distribution function fit. The error bars are errors from function fit only. The central Au + Au collision data are plotted with systematic errors. The results from HIJING and PYTHIA calculations are drawn in different line styles.

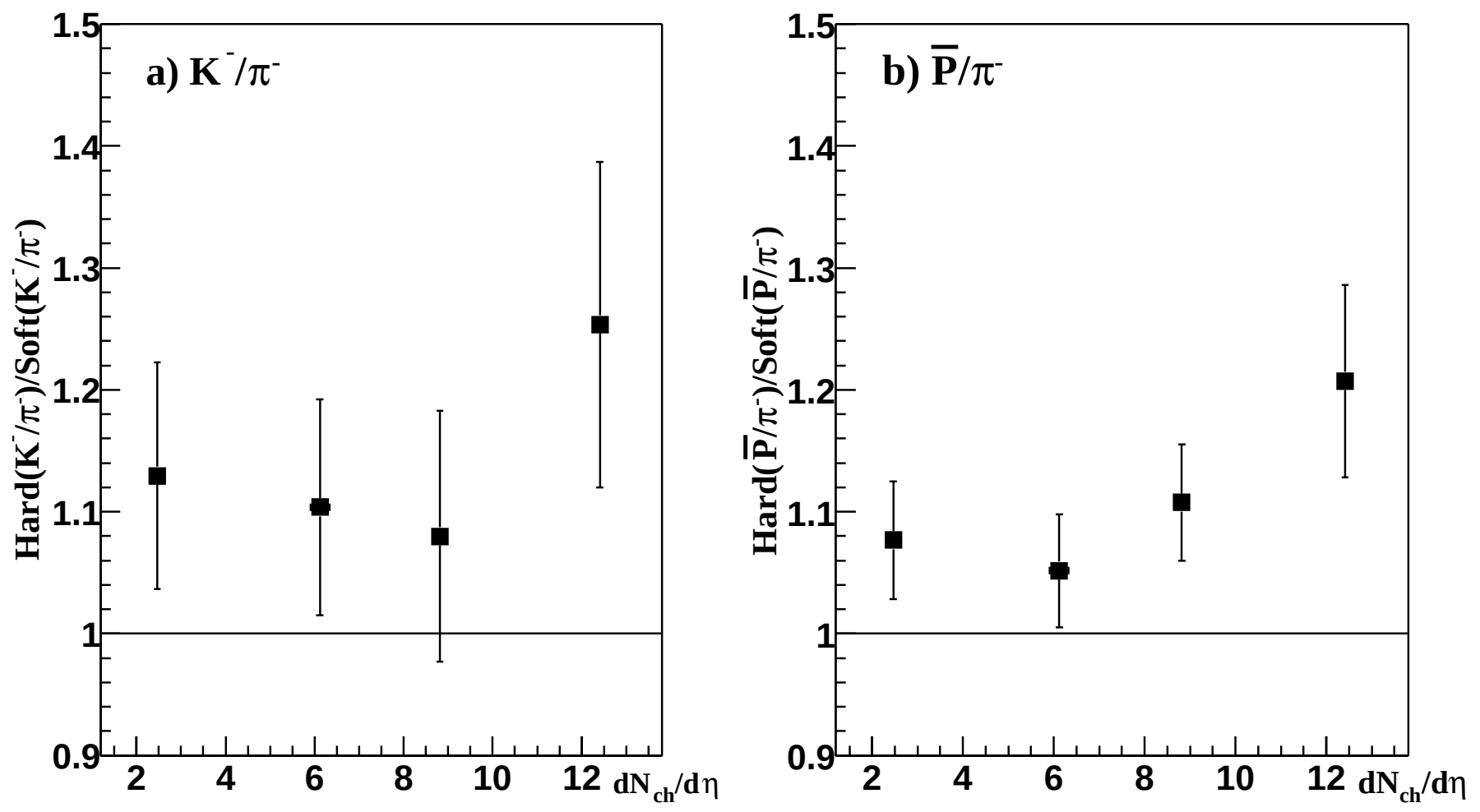

Fig4. The distributions of hard(K-/π)/soft(K-/π),hard( $\bar{P}$ /π-)/soft( $\bar{P}$ /π-) in the midrapidity (|y|<0.2). The error bars are systematic errors.

The $K^-/\pi^-$, $\bar{P}$ $/\pi^-$ ratios of the integrated *dN/dy* yields within |y|<0.2 are plotted in the Fig3. The $\pi^-$, $K^-$ and $\bar{P}$ yields are extracted by the $m_T$ exponential distribution function. The systematic errors for the $K^-/\pi^-$, $\bar{P}$ $/\pi^-$ ratios are about 7%. To compare the $K^-/\pi^-$, $\bar{P}$ $/\pi^-$ ratios in hard and soft events, we calculate the $K^-/\pi^-$, $\bar{P}$ $/\pi^-$ ratios by different fit functions to get the hard($K^-/\pi$)/soft($K^-/\pi$) and hard( $\bar{P}$ $/\pi^-$)/soft( $\bar{P}$ $/\pi^-$) as shown in the Fig4. From Fig4, we can see that $K^-/\pi^-$, $\bar{P}$ $/\pi^-$ ratios is higher in hard events than those in soft events in three multiplicity bins and these generic can be reproduced by PYTHIA[9] and HIJING calculations, which are shown in the Fig3. But both HIJING and PYTHIA calculation can't describe the multiplicity dependence of identified particle ratios. In central Au +Au collisions, the $K^-/\pi^-$, $\bar{P}$ $/\pi^-$ ratios are higher than those in soft events and are similar to those in hard events.

In summary, we have reported the results of the hard, soft and minimum bias events in 200GeV proton + proton collisions. From charged hadron <$P_T$> dependence on multiplicity and comparing with CDF results, we see that the behavior of the soft events has little dependence on collision energies, but the hard and minimum bias events are otherwise. These results confirm the conclusions from CDF Collaboration that the dynamical mechanism of inelastic multi-particle production in soft interactions from RHIC to Tevetron is energy invariant. The <$P_T$> dependence on the multiplicity of charged hadron and identified particles in soft event indicates that the mini-jet is not the only source responsible for the <$P_T$> increase with multiplicity. The $K^-/\pi^-$, $\overline{P}/\pi^-$ ratios in the hard event class are higher than those in soft event class. This generic feature is possible due to the hard or semi-hard process contribution and can be reproduced by HIJING and PYTHIA calculations, qualitatively.

**Reference:**

[1] R. Ansari *et al.*, Z. Phys. C **36**, 175 (1987); X. Wang and R.C. Hwa, Phys. Rev. D **39**,187(1989);

[2] T. Sjöstrand and M. van Zijl, Phys. Rev. D **36**, 2019 (1987).

[3] D. Acosta et al ( CDF Collaboration ). Phys. Rev. D65, 072005,(2002).

[4] C. Adler, et al. (STAR Collaboration), Phys. Rev. Lett.87, 262302 (2001).

[5] C. Adler, et al. (STAR Collaboration), Phys. Rev. Lett. 86, 4778 (2001).

[6] J. Adams et al. (STAR Collaboration), nucl-ex/0310004

[7] Xin-Nian Wang and Miklos Gyulassy, Phys. Rev. D **44**, 3501(1991)

[8] P.Abreu, et al. , (DELPHI Collaboration ), Z. Phys. C 73, 11–59 (1996)

[9] T. Sjöstrand, Comput. Phys. Commun. **82**, 74 91994); G.Marchesini, B.R. Webber, G.Abbiendi, I.G. Knowles, M.H.Seymour, and L. Stanco, *ibid.* **67**, 465 (1992).

[10] C. Adler, et al. (STAR Collaboration), Phys. Rev. Lett. 86, 4778 (2001).

[11] J. Adams et al. (STAR Collaboration), nucl-ex/0310004